\documentclass[twocolumn,prb,superscriptaddress,showpacs]{revtex4}
\newcommand{\tc}{$T_{\rm c}$}

\newcommand{\lasr}{La$_{2-x}$Sr$_x$CuO$_4$}

\newcommand{\cana}{Ca$_{2-x}$Na$_x$CuO$_2$Cl$_2$}
\newcommand{\msr}{$\mu$SR}

\usepackage{graphicx}
\usepackage{dcolumn}
\usepackage{bm}

\begin{document}


\title{Field-induced magnetism in the high-$T_c$ superconductor \cana\ \\
with flat CuO$_2$ planes
}

\author{K.~Ohishi}
\altaffiliation[Present Address: ]{Advanced Meson Science Laboratory, 
Nishina Center for Accelerator-Based Science, RIKEN, Wako 351-0198, Japan}
\affiliation{Institute of Materials Structure Science, High Energy Accelerator Research 
Organization (KEK), Tsukuba, Ibaraki 305-0801, Japan}
\author{I.~Yamada}
\altaffiliation[Present Address: ]{Graduate School of Science and Engineering, 
Ehime University, Matsuyama 790-8577, Japan}
\affiliation{Institute for Chemical Research, Kyoto University, Uji, Kyoto 611-0011, Japan}
\author{A.~Koda}
\affiliation{Institute of Materials Structure Science, High Energy Accelerator Research Organization (KEK), Tsukuba, Ibaraki 305-0801, Japan}
\affiliation{Department of Materials Structure Science, The Graduate University for Advanced Studies (SOKENDAI), Tsukuba, Ibaraki 305-0801, Japan}
\author{S.R.~Saha}
\altaffiliation[Present Address: ]{Department of Physics, University of Maryland, College Park, Maryland 20742-4111, USA}
\affiliation{Institute of Materials Structure Science, High Energy Accelerator Research Organization (KEK), Tsukuba, Ibaraki 305-0801, Japan}
\author{R.~Kadono}
\affiliation{Institute of Materials Structure Science, High Energy Accelerator Research Organization (KEK), Tsukuba, Ibaraki 305-0801, Japan}
\affiliation{Department of Materials Structure Science, The Graduate University for Advanced Studies (SOKENDAI), Tsukuba, Ibaraki 305-0801, Japan}
\author{W.~Higemoto}
\affiliation{Advanced Science Research Center, Japan Atomic Energy Agency, Tokai, 
Ibaraki 319-1195, Japan}
\author{K.M.~Kojima}
\affiliation{Department of Physics, Graduate School of Science, University of Tokyo, Bunkyo-ku, Tokyo 113-0033, Japan}
\author{M.~Azuma}
\affiliation{Institute for Chemical Research, Kyoto University, Uji, Kyoto 611-0011, Japan}
\author{M.~Takano}
\altaffiliation[Present Address: ]{Institute for Integrated Cell-Material Sciences, Kyoto 
University, Kyoto 606-8501, Japan}
\affiliation{Institute for Chemical Research, Kyoto University, Uji, Kyoto 611-0011, Japan}

\date{\today}

\begin{abstract}
The internal magnetic field distribution in a mixed state of a cuprate superconductor, \cana\ 
($T_{\rm c}\simeq28.5$ K, 
near the optimal doping), was measured by muon spin rotation (\msr) technique up to 60 kOe. The \msr\ linewidth 
$\Lambda(B)$ which exhibits excess broadening at higher fields ($B>5$ kOe) due to field-induced magnetism (FIM), 
is described by a relation, $\Lambda(B)\propto\sqrt{B}$. 
This suggests that the orbital current and associated quasiparticle excitation plays predominant roles in stabilizing 
the quasistatic correlation. Moreover, a slowing down of the vortex fluctuation sets in well above \tc\, 
as inferred from the trace of FIM observed up to $\sim80$ K, and develops continuously 
without a singularity at \tc\ as the temperature decreases.
\end{abstract}

\maketitle


The 1/8 anomaly, or the suppression of superconductivity at a specific hole carrier concentration, $p\simeq1/8$, 
has been understood in terms of the so-called ``stripe" model, where a quasistatic charge and spin stripe order develops in place of superconductivity \cite{Tranquada:95}. While a similar situation is realized over a wide range of $p\simeq x$ in La$_{2-x-y}$Nd$_y$Sr$_x$CuO$_4$ (Ref.\onlinecite{Tranquada:95}), a spatially modulated dynamical spin correlation is found in \lasr\/ (LSCO) that becomes static only near $p\simeq1/8$ (Ref.\onlinecite{Yamada:98}). It is speculated that the dynamical stripes tend to be static at $p\sim$ 1/8 by a synchronization of the periodicity between lattice and charge/spin modulations, as the buckling of CuO$_2$ planes in the tetragonal structure observed at low temperatures (LTT phase) seems to serve as a source of ``pinning", leading to a stronger 1/8 anomaly. 

Recent neutron scattering experiments in underdoped and optimally doped LSCO \cite{Katano:00,Lake:01,Lake:02} 
and underdoped La$_{2-x}$Ba$_x$CuO$_4$ (LBCO) \cite{Fujita:06} have revealed that the quasistatic stripe order is 
enhanced at around $p\simeq1/8$ by applying a moderate magnetic field of a few tesla parallel to the $c$ axis, as 
inferred from the enhanced intensity of incommensurate magnetic Bragg peaks. 
Moreover, such a field-induced static magnetic order has turned out to exhibit a three-dimensional correlation, 
which survives into the superconducting state, in LSCO ($p=0.10$), suggesting a predominant role of the 
orbital current in field-induced magnetism (FIM) \cite{Lake:05}. 

The occurrence of FIM in high-\tc\/ cuprates, which includes static and/or dynamical magnetic correlations, has 
also been suggested from muon spin rotation (\msr) measurements in underdoped and optimally doped LSCO \cite{Kadono:04,Savici:05,Ishida:07,Chang:08}, LBCO \cite{Savici:05}, La$_{2-x-y}$Eu$_y$Sr$_x$CuO$_4$ (LESCO) \cite{Savici:05}, and YBa$_2$Cu$_3$O$_{7-\delta}$ (YBCO) \cite{Sonier:08}. There is a common feature that the muon spin relaxation is observed well above \tc\ in these cuprates, suggesting that the FIM develops well above \tc. 
Recently, the FIM has been observed even in overdoped LSCO \cite{Sonier:08,MacDougall:06,MacDougall}.
The field-induced static magnetic order observed by neutron scattering experiments, on the other hand, occurs only below $T\sim$ \tc\ in underdoped and optimally doped samples of LSCO. Here, the field-induced static magnetic order is presumed to originate from quasistatic moments around the vortices, and such a magnetic order is not observed for overdoped samples. 
A lack of experimental evidence from neutrons for static magnetic order persisting above \tc\ led to the argument that the quasistatic FIM observed above \tc\ by \msr\ could be due to uncorrelated static spins. 

Regardless of such a correlation with \tc, the field-induced static magnetic order has a common feature that it is observed in those high-\tc\ cuprates that exhibit a buckling of the CuO$_2$ planes. From this view point, \cana\ (Na-CCOC) is a good candidate for studying the relation between the FIM and the buckling of CuO$_2$ planes, because Na-CCOC has flat CuO$_2$ planes without buckling, and therefore it would serve as a stage for testing the FIM as an intrinsic character of CuO$_2$ planes. 

In this paper, we demonstrate the occurrence of FIM in Na-CCOC probed by \msr. While the crystal structure of Na-CCOC is isostructural with that of LSCO, it consists of flat CuO$_2$ planes, owing to the substitution of apical oxygen with chlorine. While the \msr\ linewidth $\Lambda$ in the mixed state decreases with increasing field at low magnetic induction, as expected for a normal flux line lattice (FLL), it exhibits a turnover at around $B\simeq5$~kOe, and increases in proportion to $\sqrt{B}$, strongly suggesting that the depolarization is due to quasistatic magnetism associated with magnetic vortices whose quasiparticle density is proportional to $\sqrt{B}$ in $d$-wave superconductors \cite{Volovik:93,Nakai:04}. It is inferred from the muon Knight shift measurement that local spins of 0.15$\mu_{\rm B}$/Cu are responsible for the observed effect. Moreover, an enhancement of $\Lambda$ is observed up to 80 K under a field of 60 kOe, which is well above \tc\ (=28.5 K). 
This reveals the possibility that the fluctuation of random vortices persists over a wide range of temperature far exceeding \tc, which is in intriguing accord with the Nernst effect \cite{Xu:00,Wang:01}.

\begin{figure}[t]
\begin{center}
\rotatebox[origin=c]{0}{\includegraphics[width=0.85\columnwidth]{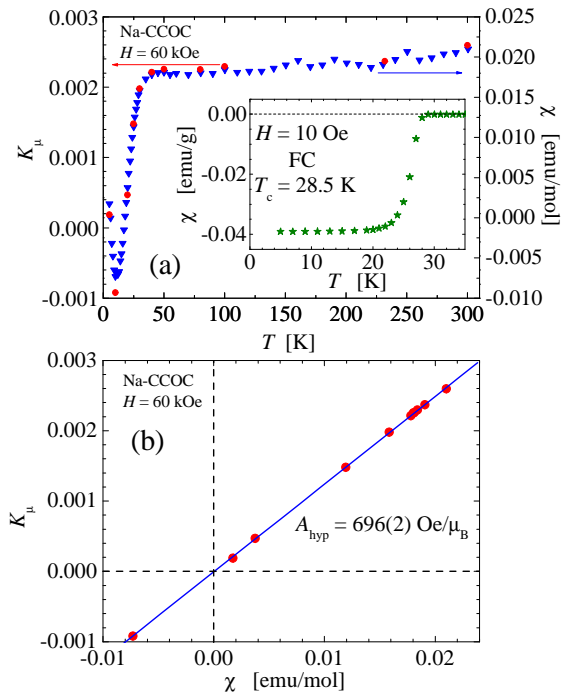}}
\caption{(Color Online) (a) Temperature dependence of the muon Knight shift, $K_\mu$ (the spin part, red circle), and susceptibility (blue triangle) at $H=60$~kOe. Inset shows the temperature dependence of susceptibility observed at $H=10$ Oe. For the origin of negative offset, see text.
(b) $K_\mu$ versus $\chi$ plot. The turnover below $\sim$10 K is attributed to unidentified impurities.}
\label{KS} 
\end{center}
\end{figure}

The single-crystalline specimen of Na-CCOC used in this study had been grown by a flux method under high pressure to yield slab samples with the $c$ axis normal to their plane \cite{Kohsaka:02,Azuma:03}. The superconducting transition temperature determined from susceptibility measurement is 28.5 K (see the inset of Fig.~\ref{KS}), which corresponds to that of optimally doped compounds according to the phase diagram \cite{Hiroi:96,Ohishi:05}. The samples were encapsulated in a polymide tape in a glove box to prevent the depletion of sodium and a subsequent deterioration. Transverse field (TF) \msr\ measurements were performed on the M15 beamline of TRIUMF, Vancouver, Canada. Muons with a momentum of 29 MeV/c were injected into the samples with their polarization rotated perpendicular to the beam momentum, so that the external field $H$ might be applied along the incoming beam axis (to minimize the disturbance to muon trajectory), which was parallel to the crystalline $c$ axis. An experimental setup with a high time resolution was employed to measure TF-\msr\ time spectra up to 60 kOe. The corresponding dc-susceptibility at 60 kOe was measured by a SQUID magnetometer. 

\begin{figure}[t]
\begin{center}
\rotatebox[origin=c]{0}{\includegraphics[width=0.85\columnwidth]{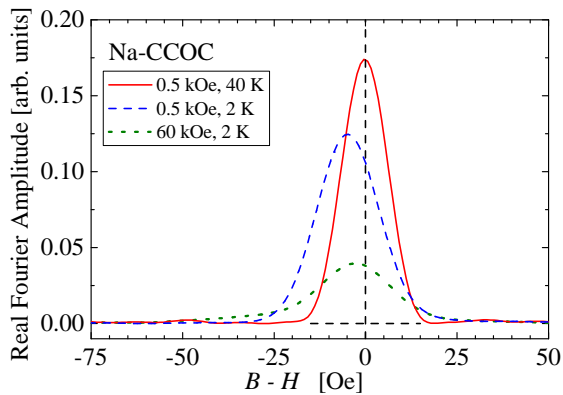}}
\caption{(Color Online) Fast Fourier transform of TF-\msr\ spectra observed at 40 K (solid curve, $H=0.5$ kOe) and 2 K (dashed curve, 0.5 kOe; dotted curve, 60 kOe).}
\label{FFT}
\end{center}
\end{figure}
Figure \ref{FFT} shows the fast Fourier transform (FFT) of TF-\msr\ time spectra, which represent the density distribution of internal magnetic field $B$ at the muon site (with additional broadening due to random local fields from nuclear magnetic moments and that coming from a finite time window of FFT). In the normal state above \tc\ (solid curve), the spectrum has a peak at $B=H=0.5$ kOe, which undergoes a shift to a lower frequency upon cooling to 2 K with an increased linewidth (dashed curve). This behavior is perfectly in line with the FLL formation in the superconducting state. Meanwhile, the spectrum exhibits further broadening with increasing field to 60 kOe (dotted curve), which is opposite to the predicted tendency of decreasing linewidth with increasing field in the conventional type II superconductors \cite{Brandt:88}. The data are analyzed by curve fits in the time domain using a phenomenological stretched exponential function, 
\begin{equation}
A\hat{P}(t) = A\exp\left[-\left(\Lambda t\right)^\beta\right]\exp(i\omega_\mu t+\phi),\label{stretch} 
\end{equation}
where $A$ is the positron decay asymmetry, $\Lambda$ is the depolarization rate
(= linewidth in the frequency domain), $\beta$ is the power of the exponent, $\omega_\mu =\gamma_\mu B$ with $\gamma_\mu$ being the muon gyromagnetic ratio ($=2\pi\times$ 13.553 MHz/kOe), and $\phi$ is the initial phase of precession. The muon Knight shift, $K$, is then defined as, 
\begin{equation}
K=\frac{\omega_\mu-\omega_0}{\omega_0}=K_\mu+K_{\rm v}+K_{\rm dem},
\label{knightshift}
\end{equation}
where $\omega_0=\gamma_\mu H$, $K_\mu$ is the shift due to the local spin susceptibility $\chi_{\rm spin}$, $K_{\rm v}$ is that due to the orbital current in the FLL state, and $K_{\rm dem}$ is the correction term consisting of demagnetization and Lorentz field [$=4\pi(1/3-N)\rho_{\rm mol}\chi_{\rm mol}$, where $N\simeq 1$ is the 
demagnetization factor, $\rho_{\rm mol}=0.01480$~mol/cm$^3$ is the molar density, 
and $\chi_{\rm mol}$ is the molar susceptibility]. 

\begin{figure}[t]
\begin{center}
\rotatebox[origin=c]{0}{\includegraphics[width=0.8\columnwidth]{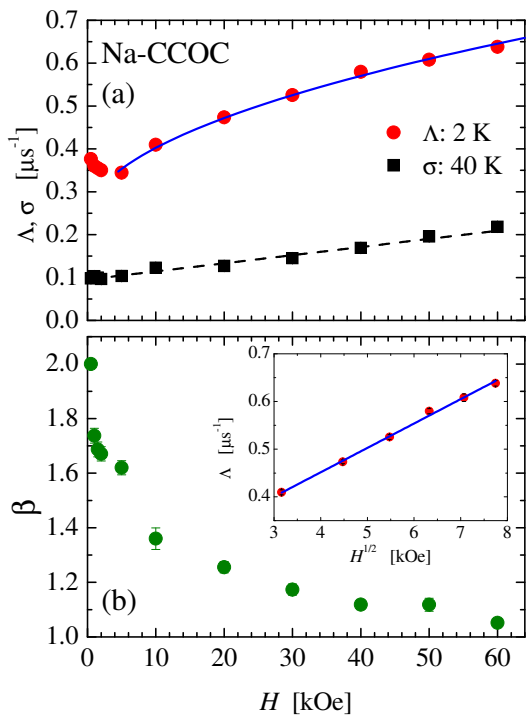}}
\caption{(Color Online) Magnetic field dependence of (a) the relaxation rate $\Lambda$ and $\sigma$ at 
2~K and 40~K, respectively, and (b) the power of exponent $\beta$ at 2~K. 
Inset shows the $\sqrt{H}$ dependence of $\Lambda$ above $H=10$~kOe. 
Solid curve represents a fit by the relation $\Lambda\propto\sqrt{H}$. }
\label{Hdep} 
\end{center}
\end{figure}
Figure~\ref{Hdep} shows the field dependence of the relaxation rate at 40 K and 2 K, and that of the power 
of the exponent at 2 K. 
The relaxation rate, $\sigma$ ($=\Lambda_{\beta=2}$), at 40 K exhibits the $H$-linear behavior, consistent with 
previous results in LSCO \cite{Savici:05,Sonier:08,MacDougall:06}, LBCO \cite{Savici:05}, LESCO \cite{Savici:05}, 
and YBCO \cite{Sonier:08}. 
Note that this linear behavior of $\sigma$ can be understood by assuming the existence of slowly fluctuating 
(staggered) random magnetic moments induced inside the vortices. As mentioned later, it is described that these 
fluctuating moments appear with the fluctuation of vortices, because the number of vortices corresponding to the 
fluctuating moments is proportional to the applied field. 
Therefore, this $H$-linear behavior of $\sigma$ is considered to be slowly fluctuating random magnetic moments. 
On the other hand, while $\Lambda$ at 2~K decreases with increasing field at lower fields, it exhibits a turnover 
around $H=5$~kOe and an increase represented by the relation $\Lambda\propto\sqrt{H}$ (where $H\simeq B$). 
The lineshape shows a change from that of the Gaussian ($\beta=2$) to single exponential decay ($\beta=1$) with 
increasing field, as shown Fig.~\ref{Hdep}(b). The contribution of FLL at low fields, $\sigma_{\rm v}$, is extracted by 
subtracting $\sigma_{\rm n}$ from the total linewidth, $\sigma$, in quadrature, 
$\sigma^2_{\rm v}=\sigma^2-\sigma^2_{\rm n}$, with $\sigma_{\rm n}$ being the depolarization rate 
in the normal state (see below). The behavior observed 
at lower fields is consistent with the predicted $H$-dependence \cite{Brandt:88},
\begin{equation}
\sqrt{2}\sigma_{\rm v}\simeq0.0274\frac{\gamma_\mu\Phi_0}{\lambda_{ab}^2(T,h)}(1-h)\sqrt{1+3.9(1-h)^2},
\label{lmdh}
\end{equation}
where $\Phi_0$ is the magnetic flux quantum, $\lambda_{ab}(T,h)$ is the effective inplane London penetration depth, which can be expressed as $\lambda_{ab}(T,0)\cdot(1-\eta h)$ with $\eta>0$ for line nodes \cite{Kadono:07}, and $h$ is the field normalized by the upper critical field ($h=H/H_{\rm c2}$). 
Meanwhile, the value of $\Lambda$ above $\sim$5 kOe far exceeds that observed at lower fields. 
It is unlikely that such an enhancement at a higher field is induced by flux pinning or other extrinsic artifacts, and thus can be uniquely attributed to FIM that is spatially inhomogeneous (as inferred from the single exponential-like lineshape). More interestingly, $\Lambda$ is excellently reproduced by the relation $\Lambda(B)\propto\sqrt{B}$ over the relevant field range. The inset of Fig.~\ref{Hdep}(b) shows $\Lambda$, which is plotted against $\sqrt{B}$ to see the linearity. 
Note that this behavior is not observed in other high-\tc\/ cuprates that do not have flat CuO$_2$ planes \cite{Savici:05,Ishida:07,Chang:08,Sonier:08,MacDougall:06,MacDougall}. 
It is established that quasiparticle excitation in the mixed state of $d$-wave superconductors is extended along the $(\pi,\pi)$ directions, leading to a non-linear field dependence of the quasiparticle density that is well approximated by such $\sqrt{B}$-dependence \cite{Volovik:93,Nakai:04}. 
Here, we would like to consider the relation between the muon spin relaxation and 
the quasiparticle excitations in $d$-wave superconductors. In the case of the {\it usual} type II superconductors with 
$d$-wave symmetry, which do not show any FIM, muon spin relaxation rate decreases with increasing $H$. 
This is because 
the induced quasiparticle excitations around the nodal region due to pair breaking makes $\lambda$ increase. 
According to a relation of Eq.~(\ref{lmdh}), $\sigma_{\rm v}$ decreases with increasing $\lambda$. 
On the other hand, in the case of superconductors, which show FIM, muon spin relaxation rate increases 
because relaxation due to magnetic moments is dominant compared with that due to FLL (increase of $\lambda$)
\cite{Kadono:04,Savici:05,Ishida:07,Chang:08,Sonier:08,MacDougall:06,MacDougall}. It is suggested that 
the induced quasiparticle excitations have quasistatic magnetic moments to describe the increase of $\Lambda(B)$. 
Consequently, we can assume that the observed field dependence of $\Lambda$ is one type of evidence for the 
$d$-wave superconductor, because quasiparticle excitations in $d$-wave superconductors shows a 
$\sqrt{B}$-dependence. 

\begin{figure}[t]
\begin{center}
\rotatebox[origin=c]{0}{\includegraphics[width=0.8\columnwidth]{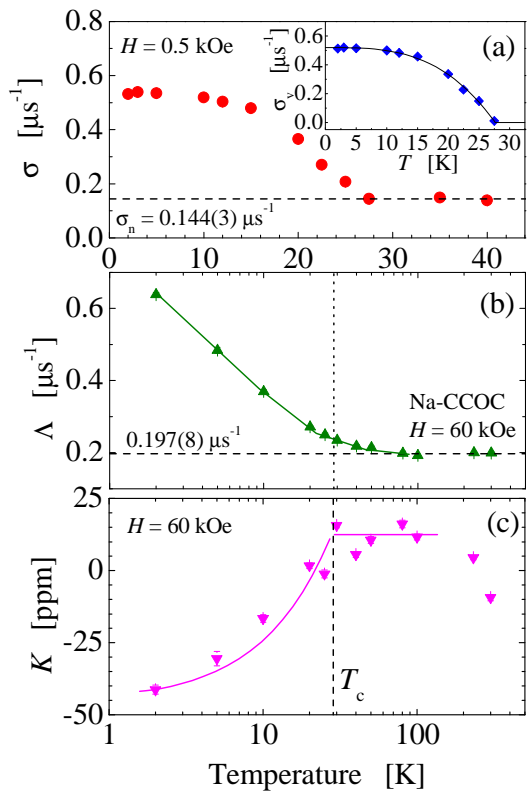}}
\caption{(Color Online) Temperature dependence of the relaxation rate (a) $\sigma$ and $\sigma_{\rm v}$ 
(inset) at $H=0.5$~kOe, (b) $\Lambda$ at $H=60$~kOe, and (c) the total muon Knight shift $K$ at $H=60$~kOe. 
The solid curve in the inset of (a) is a fit with $\sigma_{\rm v}(T)=\sigma(0)[1-(T/T_{\rm c})^n]$. Those in (b) and (c) are the guides for the eyes.}
\label{rlx-T}
\end{center}
\end{figure}
As shown in Fig.~\ref{rlx-T}(a), while the linewidth, $\sigma$ ($=\Lambda_{\beta=2}$ observed at 0.5 kOe), is mostly independent of temperature above \tc\ with a mean value of $\sigma_{\rm n}=0.144(3)$ $\mu$s$^{-1}$, it increases with decreasing temperature below \tc\ as
FLL is formed. The value of $\sigma_{\rm n}$ is in perfect agreement with that of random local fields from nuclear moments estimated from a previous \msr\ experiment \cite{Ohishi:05}. 
Fits by a power law, $\sigma_{\rm v}(T)=\sigma_{\rm v}(0)\left[1-\left(T/T_{\rm c}\right)^n\right]$, with \tc\/ as a free parameter yields $\sigma_{\rm v}(0)=0.519(5)$ $\mu$s$^{-1}$, $T_{\rm c}=27.7(2)$~K, and $n=3.2(1)$ [Fig.~\ref{rlx-T}(a), inset]. Using Eq.~(\ref{lmdh}) for $h\ll1$, the magnetic penetration depth, $\lambda_{ab}(0,0)$, extrapolated to $T=0$~K is evaluated to be 382(4) nm. These results are quantitatively consistent with earlier literature on Na-CCOC (with the sodium concentration $x=0.18$, and $H=2$~kOe), except that no anomalous increase of $\sigma$ was observed below $\sim$5~K \cite{Khasanov:07}.

The temperature dependence of $\Lambda$ observed at $H=60$~kOe is shown in Fig.~\ref{rlx-T}(b). 
It increases below 80~K ($\gg T_c$), and continues to increase down to the lowest temperature 
without any saturation. This behavior is similar to those observed in LSCO at a high field 
\cite{Savici:05,Ishida:07,Chang:08,Sonier:08,MacDougall:06}. 
On the other hand, the muon Knight shift, $K$, decreases with decreasing temperature below 
\tc\ [Fig.~\ref{rlx-T}(c)], indicating that both the FIM and superconductivity coexist below \tc. 
The behavior of $K$ is in line with that of the magnetic susceptibility [$\chi$, measured after cooling under 
a field of 60 kOe, shown in Fig.~\ref{KS}(a)], except that below $\sim$10 K (see below), 
although the contributions of spin and orbital parts are not separated in these quantities. 
The spin part of the muon Knight shift, $K_\mu$, is extracted by Eq.~(\ref{knightshift}) for 
$T>T_{\rm c}$ (where $K_{\rm v}=0$) and mapped into the $K_\mu$-$\chi$ plot [Fig.~\ref{KS}(b)]. 
All of the data points fall on a straight line without any sign of kink for those obtained below $\sim$80 K (smaller values of $K_\mu$-$\chi$), indicating that FIM is indeed carried by the paramagnetic Cu spins. 
The muon hyperfine parameter, $A_\mu$, is deduced as a gradient, $dK_\mu/d\chi$, 
in the normal state according to, 
\begin{equation}
K_\mu=K_0+\frac1{N_{\rm A}\mu_{\rm B}}A_\mu\chi,
\end{equation}
which yields $A_\mu=696(2)$~Oe/$\mu_{\rm B}$,
where $K_0$ ($\simeq0$) is a $T$-independent Fermi contact coupling with the conduction electrons, 
$N_{\rm A}$ is the Avogadro's number and $\mu_{\rm B}$ is the Bohr magneton. 
The mean value $|\overline{\mu}_{\rm Cu}|$ for the moment size of the local copper spins is then obtained by comparing $A_\mu$ with that determined by the $z$-component of the dipole tensor,
\begin{equation}
\frac{|\overline{\mu}_{\rm Cu}|}{\mu_{\rm B}}\sum_k a_k A^{zz}_{k}=
\frac{|\overline{\mu}_{\rm Cu}|}{\mu_{\rm B}}\sum_k a_k\sum_i\frac{1}{r^3_{ik}}\left[\frac{3z^2_{ik}}{r^2_{ik}}-1\right]\simeq A_\mu, 
\label{dtensor}
\end{equation}
where $a_k$ is the relative population of the $k$-th muon site 
(there are three of them in Na-CCOC \cite{Ohishi:05}, see below), $r_{ik}$ is the distance 
between muon at the $k$-th site and the $i$-th Cu ions (with $z_{ik}$ being their $z$-component). Combining 
$A_\mu$ with other information on the local magnetic fields at the muon sites previously obtained for 
antiferromagnetically ordered phase of the parent compounds \cite{Ohishi:05}, we reexamined the muon sites to reproduce all of the quantities concerning the local fields. The result is summarized in Table~\ref{t1}, 
where the $\mu_1$ site turns out to be slightly off the previously assigned position as inferred 
from the sign of $A_\mu$. Accordingly, we have 
$\overline{A}^{zz}_{\rm dip}=\sum_k a_k A^{zz}_k=4788$~Oe$/\mu_{\rm B}$, and the effective moment size 
$|\overline{\mu}_{\rm Cu}|=0.1454(4)\mu_{\rm B}$ from Eq.~(\ref{dtensor}). 

\begin{table}[tb]
\caption{Summary of parameters for muon sites identified in Na-CCOC, where 
$a_k$ is the fractional yield, $r_{\rm Cu}$ is the distance between muon, 
$A^{zz}_k$ is the dipole tensor calculated for the muon coordinates shown in the preceding column. 
The crystal structure of Na-CCOC is K$_2$NiF$_4$-type structure (space group: $I4/mmm$) with lattice constants 
$a=3.8687$\AA\/ and $c=15.0485$\AA.}
\label{t1}
\begin{tabular}{c|cc|cc}
\hline\hline
$\mu_k$ site & $a_k$ & $r_{\rm Cu}$ [nm] & Coordinates & $A^{zz}_k$ [Oe/$\mu_{\rm B}$]\\ 
\hline\hline
$\mu_1$ & 0.68(2)$^a$ & 0.138$^a$ & (0,0,0.0917) & 6678 \\
$\mu_2$ & 0.08(1)$^a$ & 0.254$^a$ & (0,0.310,0.146) & 813 \\
$\mu_3$ & 0.23(1)$^a$ & 0.343$^a$ & (0,0.5,0.186) & 637 \\
\hline\hline
\multicolumn{5}{l}
{\small $a$: Ref.~\cite{Ohishi:05}}\\
\end{tabular}
\end{table}

It is noteworthy in Fig.~\ref{KS}(a) that $\chi$ exhibits an upturn below $\sim$10 K, which might hint the occurrence of magnetic order. However, no such behavior is observed in the muon Knight shift $K$ shown in Fig.~\ref{rlx-T}(c); it must be noted that $K_\mu$ (the spin part) is deduced using Eq.~(\ref{knightshift}) where $\chi$ comes in via the demagnetization term ($K_{\rm dem}\propto\chi$), so that the temperature dependence of $K_\mu$ is mostly parallel with that of $\chi$. We attribute this upturn in $\chi$ to unidentified impurities at this stage, considering the absence of a corresponding anomaly in $A_\mu$ anticipated for quasistatic magnetic order. 
The ambiguity on the absolute values of $\chi$ makes it difficult to discuss whether or not the observed change of sign in $K_\mu$ below $\sim$20 K is meaningful. However, 
such a behavior is readily explained by considering the contribution of $K_{\rm v}$ ($<0$) which is not discernible from other contributions in the present experiment.

Since Na-CCOC does not exhibit a quasistatic stripe correlation under a zero external field over 
the relevant hole concentration \cite{Ohishi:05}, the occurrence of FIM can not be explained by 
a residual effect of the 1/8 anomaly, as it might have been in LSCO \cite{Savici:02}. 
A recent revelation of a weak 1/8 anomaly observed in Na-CCOC upon Zn substitution for Cu \cite{Satoh} strongly suggests that the situation is similar to that observed in YBCO \cite{Akoshima:00} and Bi$_2$Sr$_2$Ca$_{1-x}$Y$_x$Cu$_2$O$_{8+\delta}$ \cite{Watanabe:00}. This, together with the present result, supports the presumption that the instability of dynamical stripe correlation against local suppression of superconducting order parameter (i.e., by the Zn impurity or flux lines) is an intrinsic property of CuO$_2$ planes, common at least in Na-CCOC, LSCO and LBCO regardless of the buckling of the CuO$_2$ planes. 

According to the reported result on the Nernst effect \cite{Xu:00,Wang:01}, it is suggested that strong fluctuation of superconductivity and associated flux lines persists well above \tc. This naturally leads to a possibility that the FIM observed above \tc\ may originate from the quasistatic stripe correlation segregated around the fluctuating vortices. 
Considering the result of field dependence of $\Lambda$ at 40~K, which behaves $H$-linear, this scenario is apparently against the expected field dependence in the FLL state \cite{Brandt:88}. However, it may be argued that the $\sqrt{B}$-dependence is expected only when the quasistatic FLL is established \cite{Volovik:93}. 
Moreover, this scenario is also supported by thermal conductivity measurements in LSCO, where the suppression of thermal conductivity observed above \tc\ is attributed to the development of the quasistatic stripe order associated with fluctuating vortices \cite{Kudo:04}. Thus, the appearance of FIM above \tc\ can be explained in terms of the instability of dynamical stripes upon the local suppression of the order parameter. 

Finally, it would be worth mentioning that the suppression of the FIM near $p\sim 1/8$ observed in YBCO \cite{Sonier:08} has little or no direct relevance to our argument. 
While the relaxation rate is suppressed near $p\sim 1/8$ compared to other hole concentrations, they indeed observe that the FIM sets in at around $210$~K at $p\sim 1/8$, which is much higher than \tc\cite{Sonier:08}. Thus, YBCO seems to share the feature of FIM that develops well above \tc\ (while the onset temperature for the FIM might exhibit the $p$-dependence similar to \tc\ characterized by a dip near $p\sim 1/8$).

In conclusion, our \msr\ measurements on optimally doped Na-CCOC demonstrate 
the occurrence of FIM above $\sim$5 kOe, which is highly inhomogeneous and coexists microscopically with superconductivity. Muon Knight shift measurements indicate that the magnetism is carried by local copper spins with an average moment size of 0.15$\mu_{\rm B}$. The field-dependence of the muon spin depolarization rate suggests that the magnetism comes from the quasistatic stripe correlation around the vortex cores (with $d$-wave paring). This is also in line with the interpretation that the observed FIM up to 80 K (well above \tc) comes from the strong fluctuation of vortices suggested by the Nernst effect. 

We would like to thank the staff of TRIUMF for their technical support during the 
experiments, and acknowledge helpful discussions with Y. Koike. 
This work was partially supported by a Grant-in-Aid for Creative Scientific 
Research (Grant No. 13NP0201) and a Grant-in-Aid for Scientific Research by the 
Ministry of Education, Culture, Sports, Science and Technology of Japan 
(Grants No. 17105002, 19052008 and 19340098).

\end{document}